\documentclass[preprint,nofootinbib,aps,superscriptaddress,showpacs]{revtex4}
\usepackage{epsfig}
\def\be{\begin{equation}} \def\ee{\end{equation}} \def\bea{\begin{eqnarray}}
\def\eea{\end{eqnarray}} \def\nnb{\nonumber}

\begin{document}
\title{Spin Polarization in $\gamma d \to \vec{n}p$ at Low Energies
with a Pionless Effective Field Theory}
\author{S.-I. Ando}
\affiliation{Department of Physics Education, Daegu University,
Gyeongsan 712-714, Republic of Korea}
\author{Y.-H. Song}
\affiliation{Department of Physics and Astronomy, 
University of South Carolina, Columbia, SC 29208, USA}
\author{C. H. Hyun}
\affiliation{Department of Physics Education, Daegu University,
Gyeongsan 712-714, Republic of Korea}
\author{K. Kubodera}
\affiliation{Department of Physics and Astronomy, 
University of South Carolina, Columbia, SC 29208, USA}

\date{May 14, 2011}

\begin{abstract}
With the use of pionless effective field theory including dibaryon fields,
we study the $\gamma d \to \vec{n} p$ reaction
for the laboratory photon energy $E_\gamma^{lab}$ ranging from threshold 
to 30 MeV.
Our main goal is to calculate the neutron polarization $P_{y'}$
defined as 
$P_{y'} = (\sigma_+ - \sigma_-)/(\sigma_+ + \sigma_-)$, where 
$\sigma_+$ and $\sigma_-$ are the differential cross sections 
for the spin-up and spin-down neutrons, respectively, 
along the axis perpendicular to the reaction plane.
We also calculate the total cross section 
as well as the differential cross section $\sigma(\theta)$,
where $\theta$ is the colatitude angle.
Although the results for the total and differential 
cross sections are found to agree reasonably well with the data, 
the results for $P_{y'}$ show significant discrepancy 
with the experiment.  We comment on this discrepancy.
\end{abstract}

\pacs{13.40.-f, 21.45.Bc, 24.70.+s}

\maketitle

\section{Introduction}
\label{section1}

The induced neutron-spin polarization $P_{y'}$
in  the  $\gamma d$ $\to$ $\vec{n} p$ reaction is defined as 
$P_{y'} = (\sigma_+ - \sigma_-)/(\sigma_+ + \sigma_-)$, 
where 
$\sigma_+$ and $\sigma_-$ are the differential cross sections 
for the spin-up and spin-down neutron, respectively,
along the axis perpendicular to the reaction plane.
Conspicuous discrepancy between the experimental and theoretical values
of $P_{y'}$ is a long-standing puzzle in low-energy nuclear physics
\cite{schi-prc05,KUKetal}.
Schiavilla~\cite{schi-prc05} carried out 
an elaborate calculation of $P_{y'}$
based on the so-called standard nuclear physics approach (SNPA).
In SNPA, the nuclear wave functions are generated 
with the use of high-precision phenomenological nucleon-nucleon potentials
that accurately reproduce thousands of neutron-proton
and proton-proton scattering data 
(for laboratory energies below 350~MeV)
along with the deuteron properties.
The electromagnetic current operators in SNPA are constructed from  
meson-exchange diagrams in such a manner that gauge invariance
and the low-energy theorems are satisfied.  
SNPA has been used to calculate a great many electromagnetic 
observables involving lightest nuclei, 
and its general quantitative success is well 
known~\cite{mvskr-prc05,mrs-prc98}.
As for the $\gamma d$ $\to$ $\vec{n} p$ reaction,
the differential cross sections calculated 
by Schiavilla~\cite{schi-prc05} 
for the lab-system photon energy, $E_\gamma^{lab}$, 
up to 30 MeV agree very well with the data.  
For the spin polarization $P_{y'}$, however,
there is a large discrepancy between the
state-of-the-art SNPA calculation~\cite{schi-prc05}
and the data.
This discrepancy gives a further indication
of the seriousness of the ``$P_{y'}$ puzzle",
and one is led to ask whether 
the problem lies with theory or experiment.

In order to shed more light on this issue,  
we study here the $\gamma d$ $\to$ $\vec{n} p$ reaction
in the framework of effective field theory (EFT).
The application of EFT to nuclear electroweak processes,
pioneered in Refs.~\cite{Rho,PMR,pmr-npa96},
has made great progress since then 
with various specific approaches and techniques
developed along the way.
It is to be emphasized that,
insofar as all the relevant low-energy constants (LECs) are known, 
EFT can give model-independent results, 
and that the accuracy of these results
can be systematically assessed
in virtue of the well-defined EFT expansion scheme.
In the present work we study the 
$\gamma d$ $\to$ $\vec{n} p$ reaction
in the framework of pionless EFT 
with dibaryon fields~\cite{k-npb97,bs-npa01,ando-prc05},
which has shown good convergence behavior in perturbative
expansion for a number of low-energy processes
in the two-nucleon systems
\cite{ando-prc06,ando-prc07,ando-plb08}.
We compare the results of our EFT calculation
with the experimental data, and also
with the theoretical results obtained in SNPA~\cite{schi-prc05}.
It is hoped that the present study will provide 
useful information regarding the $P_{y'}$ puzzle.

This paper is organized as follows. 
We describe in Sec.~\ref{section2}
the basic elements such as Lagrangians,
%
%
%
the definitions of observables, and the electromagnetic
operators. 
In Sec.~\ref{section3} we enumerate Feynman diagrams that appear
up to the order under consideration, 
and evaluate their amplitudes.
Sec.~\ref{section4} explains the relation between 
the amplitudes and observables.
In Sec.~\ref{section5} we show the numerical results and 
compare them with data as well as with the results of
the previous theoretical work.
Sec.~\ref{section6} is dedicated to conclusions.

\section{Formalism}
\label{section2}

As stated, we work in the framework of pionless EFT 
with dibaryons (dEFT for short).  
Since we follow the same formalism as in Ref.~\cite{ando-prc05},
we only give its brief summary here,
relegating details to Ref.~\cite{ando-prc05}.
We consider two dibaryons,
one in the $^1S_0$ channel and the other in the $^3S_1$ channel,
and denote them by $d^s$ and $d^t$, respectively.
A dEFT Lagrangian for a case involving  
an external vector field is given by
\bea
{\cal L}_{\rm dEFT} &=&
{\cal L}_N
+ {\cal L}_s
+ {\cal L}_t
+ {\cal L}_{st}\,,\label{eq:LdEFT}
\eea
where ${\cal L}_N$ is the standard heavy-nucleon Lagrangian 
for the one-nucleon sector;
${\cal L}_s$ (${\cal L}_t$) is a Lagrangian 
for $d^s$ ($d^t$), while 
${\cal L}_{st}$ describes $d^s$-$d^t$ transition
due to an external vector field (a photon field).  
We employ the standard counting rules and calculate 
the amplitude up to next-to leading order (NLO).
The ${\cal L}_N$ relevant to our NLO calculation reads
\bea
{\cal L}_N &=& N^\dagger \left[
iv\cdot D
+ \frac{1}{2m_N} \left[
(v\cdot D)^2 -D^2 \right]
-i[S^\mu,S^\nu] \left(
\mu_V f^+_{\mu\nu}
+\mu_S v^S_{\mu\nu}\right)
\right] N\, ,
\eea
where $v^\mu$ is a velocity vector satisfying $v^2=1$,
and $S^\mu$ is the nucleon spin operator. 
Here we choose $v^\mu=(1,\vec{0})$ and, correspondingly, 
$2S^\mu=(0,\vec{\sigma})$.
$D_\mu = \partial_\mu - \frac{i}{2} \vec{\tau}\cdot \vec{\cal V}_\mu 
- \frac{i}{2} {\cal V}^S_\mu
=\partial_\mu-iQ V_\mu^{\rm ext}$ is the covariant derivative 
involving the isoscalar-vector and 
isovector-vector fields, $Q$ is an electric charge of a nucleon
and $f^+_{\mu\nu}$ and $v^S_{\mu\nu}$ are the 
isovector and isoscalar field strength tensors, respectively.
$m_N$ is the nucleon mass, while $\mu_V=4.706$ and $\mu_S=0.880$ are 
the isovector and isoscalar 
magnetic moments of 
the nucleon, respectively.

The dibaryon Lagrangians, ${\cal L}_s$ and ${\cal L}_t$, 
and the transition Lagrangian, ${\cal L}_{st}$, are given by
\bea
{\cal L}_s &=& 
-s_i^\dagger \left[
iv\cdot {\cal D}
+\frac{1}{4m_N}[(v\cdot {\cal D})^2 - {\cal D}^2]
+\Delta_s\right] s_i
-y_s \left[
s_i^\dagger (N^TP_i^{(^1S_0)}N) + \mbox{\rm h.c.}
\right],
\label{eq:Ls}
\\
{\cal L}_t &=& 
-t_i^\dagger \left[
iv\cdot {\cal D}
+ \frac{1}{4m_N} [(v\cdot {\cal D})^2- {\cal D}^2]
+\Delta_t \right] t_i
-y_t \left[
t_i^\dagger (N^TP_i^{(^3S_1)}N) + \mbox{\rm h.c.} \right]
\nnb \\ 
&& -\frac{2L_2}{m_N\rho_d} i\epsilon_{ijk}t^\dagger_i t_j B_k,
\label{eq:Lt}
\\
{\cal L}_{st} &=& \frac{L_1}{m_N\sqrt{r_0\rho_d}}
\left[
t_i^\dagger s_3 B_i + \mbox{\rm h.c.}
\right]\,,\label{eq:Lst}
\eea
${\cal D}_\mu = \partial_\mu -i C V^{\rm ext}_\mu$ 
is the covariant derivative coupled with the external
vector field, where $C$ is the electric charge of 
the dibaryon field in units of the proton charge;  
$C$ = 2, 1, and 0 for the $pp$-, $np$- and 
$nn$-channel dibaryons, respectively.
$\Delta_s$ ($\Delta_t$) is the difference between 
the mass of $d_s$ ($d_t$) and $2m_N$.
$y_s$ ($y_t$) specifies the strength
of $d_s$-$N$-$N$ ($d_t$-$N$-$N$) coupling.
$P^{(^1S_0)}_a$ and $P^{(^3S_1)}_i$ are projectors 
onto the $^1S_0$ and $^3S_1$ two-nucleon states, respectively: 
\bea
P^{(^1S_0)}_a = \frac{1}{\sqrt8}\tau_2\tau_a\sigma_2\,,
\ \ \ 
P^{(^3S_1)}_i = \frac{1}{\sqrt8}\tau_2\sigma_2\sigma_i\,,
\eea
where $\tau_a$ and $\sigma_i$ are the isospin and spin operators.
$\vec{B}$ in Eqs.~(\ref{eq:Ls}) and (\ref{eq:Lst}) 
is the magnetic field,
$\vec{B}= \vec{\nabla}\times \vec{V}^{\rm ext}$,
where $\vec{V}^{\rm ext}$ is the external vector field.
$L_1$ is a 
LEC representing the strength 
of a $\vec{V}^{\rm ext}$-$d^s$-$d^t$ vertex; 
$L_2$ is a second LEC parameterizing the strength of
a $\vec{V}^{\rm ext}$-$d^t$-$d^t$ vertex.  
$\rho_d$ and $r_0$ are the effective range parameters 
of the $NN$ interaction for the deuteron and spin-singlet channel, 
respectively.

The parameters, $\Delta_{s,t}$ and $y_{s,t}$,
in Eqs.~(\ref{eq:Ls}) and (\ref{eq:Lt})
can be fixed from the scattering length 
and effective range for the $^1S_0$ and $^3S_1$ channels.
Meanwhile, $L_1$ and $L_2$ can be determined
from the low-energy $np\to d\gamma$
cross section and the deuteron magnetic moment, respectively.
Hence there are no unknown parameters 
in the Lagrangian, ${\cal L}_{\rm dEFT}$, in Eq.~(\ref{eq:LdEFT});
see Ref.\cite{ando-prc05} for further details.

\section{Transition Amplitudes}
\label{section3}

\begin{figure}[t]
\begin{center}
\epsfig{file=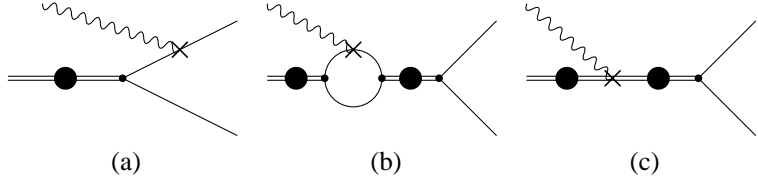,width=10.0cm}
\caption{\it Diagrams for $d\gamma\to np$ reaction:
a double line with a filled circle stands for 
a dressed dibaryon field, 
a single line for a nucleon, and a wavy line for a photon field.
The photon-nucleon-nucleon vertex with ``$\times$" is of NLO,
and the photon-dibaryon-dibaryon vertex with ``$\times$"
is proportional to the LEC, $L_1$ or $L_2$.   
\label{fig;diagrams-dgnp}}
\end{center}
\end{figure}

The Feynman diagrams contributing to our NLO calculation
are depicted in Fig.~\ref{fig;diagrams-dgnp}.
The transition amplitude $A$, 
in the center-of-mass (c.m.) frame, 
corresponding to diagrams (a), (b), and (c)
in Fig.~\ref{fig;diagrams-dgnp} may be written as
\bea
A &=&
\chi_1^\dagger \vec{\sigma}\sigma_2\tau_2\chi_2^{T\dagger}\cdot
\left\{
[\vec{\epsilon}_{(d)}\times(\hat{k}\times\vec{\epsilon}_{(\gamma)})] X_{MS}
+ \vec{\epsilon}_{(d)}\vec{\epsilon}_{(\gamma)}\cdot\hat{p}\, Y_{ES}
\right\}
\nnb \\ &&
+ \chi_1^\dagger \sigma_2\tau_3\tau_2\chi_2^{T\dagger}
i\vec{\epsilon}_{(d)}\cdot (\hat{k}\times\vec{\epsilon}_{(\gamma)})\,
X_{MV}
\nnb \\ &&
+ \chi_1^\dagger \vec{\sigma}\sigma_2\tau_3\tau_2\chi_2^{T\dagger} \cdot
\left\{
\vec{\epsilon}_{(d)}\vec{\epsilon}_{(\gamma)}\cdot\hat{p}\, X_{EV}
+ [\vec{\epsilon}_{(d)}\times(\hat{k}\times\vec{\epsilon}_{(\gamma)})]\,
Y_{MV}
\right\}
\nnb \\ &&
+ \chi_1^\dagger \sigma_2\tau_2\chi_2^{T\dagger} \,
i\vec{\epsilon}_{(d)}\cdot (\hat{k}\times \vec{\epsilon}_{(\gamma)})\,
Y_{MS}\,,\label{eq:Amplitude}
\eea
where
$\vec{\epsilon}_{(d)}$ and $\vec{\epsilon}_{(\gamma)}$ are 
spin polarization vectors for the incoming deuteron and photon,
respectively, while $\chi_1^\dagger$ and $\chi_2^\dagger$ are 
the spinors of the outgoing nucleons.
$\vec{k}$ is the momentum of an incoming photon
(which is taken to be in the $z$-direction), 
and $\vec{p}$ is the relative three-momentum of 
the two nucleons in the final state,
and we have introduced
$\hat{k}\equiv\vec{k}/|\vec{k}|$ 
and $\hat{p}\equiv\vec{p}/|\vec{p}|$.
The coefficients of the terms in Eq.~(\ref{eq:Amplitude})
are given as 
\bea
X_{MV} &=& - \sqrt{\frac{\pi\gamma}{1-\gamma\rho_d}}
\frac{1}{
\frac{1}{a_0}
+ip
-\frac12r_0p^2}
\frac{1}{2m_N}
\nnb \\ && \times
\left\{
\mu_V\left[
{\rm arccos}\left(
\frac{m_N}{\sqrt{
(m_N+\frac12\omega)^2-p^2
}}
\right)
+i\ln\left(
\frac{m_N+\frac12\omega+p}{\sqrt{
(m_N+\frac12\omega)^2-p^2
}}
\right)
\right]
\right.
\nnb \\ && \left.
- \frac{\mu_V}{m_N} \left(
\frac{1}{a_0}
+ ip
-\frac12r_0 p^2
\right) F^+
+ \omega L_1
\right\}\,,
\\
X_{MS} &=& - \sqrt{\frac{\pi\gamma}{1-\gamma\rho_d}}
\frac{1}{
\gamma
+ip
-\frac12\rho_d(\gamma^2+p^2)}
\frac{1}{2m_N}
\nnb \\ && \times
\left\{
\mu_S\left[
{\rm arccos}\left(
\frac{m_N}{\sqrt{
(m_N+\frac12\omega)^2-p^2
}}
\right)
+i\ln\left(
\frac{m_N+\frac12\omega+p}{\sqrt{
(m_N+\frac12\omega)^2-p^2
}}
\right)
\right]
\right.
\nnb \\ && \left.
- \frac{\mu_S}{m_N} \left[
\gamma 
+ ip
-\frac12\rho_d(\gamma^2+p^2)
\right] F^+
+ 2 \omega L_2
\right\}\,,
\\
X_{EV} &=& \sqrt{
\frac{\pi\gamma}{1-\gamma\rho_d}
} \frac{1}{m_N^2}
\frac{p}{\omega}F^+\,,
\ \ \
Y_{ES} = \sqrt{
\frac{\pi\gamma}{1-\gamma\rho_d}
} \frac{1}{m_N^2}
\frac{p}{\omega}F^-\,,
\\
Y_{MV} &=& \sqrt{
\frac{\pi\gamma}{1-\gamma\rho_d}
} \frac{\mu_V}{2m_N^2} F^-\,,
\ \ \
Y_{MS} = \sqrt{
\frac{\pi\gamma}{1-\gamma\rho_d}
} \frac{\mu_S}{2m_N^2} F^-\,.
\eea
with
\bea
2F^\pm =
\frac{1}{1 + \frac{\omega}{2m_N}
- \frac{\vec{p}\cdot\hat{k}}{m_N}}
\pm \frac{1}{1 + \frac{\omega}{2m_N}
+ \frac{\vec{p}\cdot\hat{k}}{m_N}}\,,
\eea
where $p=|\vec{p}|$, 
and $\omega$ is the incoming photon energy
in the c.m. frame. 

\section{Differential cross section and neutron spin polarization}
\label{section4}

We calculate the differential cross section 
and neutron spin polarization $P_{y'}$
in the c.m.\ frame.\footnote{
If comparison with the experimental data necessitates it,
we shall convert them into laboratory-frame quantities.
Numerically, this conversion is not important 
in the present case.} 
The differential cross section is given as
\bea
\sigma(\theta) =
\frac{d\sigma}{d\Omega}
= \frac{\alpha}{24\pi}\frac{pE_1}{\omega}\sum_{spin}|A|^2\,,
\label{eq;dsig}
\eea 
where
\bea
S^{-1}\sum_{spin} |A|^2 &=&
16\left(
|X_{MS}|^2 + |Y_{MV}|^2
\right)
+ 8\left(
|X_{MV}|^2 + |Y_{MS}|^2
\right)
\nnb \\ &&
+ 12 [1-(\hat{p}\cdot\hat{k})^2]
\left(
|X_{EV}|^2 + |Y_{ES}|^2
\right) \,.
\eea
The symmetry factor $S$ is equal to 2 in the present case. 
In Eq.~(\ref{eq;dsig}),
$\alpha$ is the fine structure constant,
$E_1=\sqrt{m_N^2+p^2}$ is the energy of an outgoing nucleon
in the c.m. frame,
and 
\bea
p = \frac12\sqrt{
(\omega+
\sqrt{m_d^2+\omega^2})^2
-4m_N^2
}\,,
\eea
where $m_d$ is the mass of the deuteron.
The total cross section is obtained 
by integrating Eq.~(\ref{eq;dsig}) over the direction of
$\vec{p}$.

To calculate the neutron spin polarization, 
we introduce the spin-isospin projection operator,
\bea
P_\pm = \frac12(1-\tau_3) 
\frac12(1\pm \vec{\sigma}\cdot \hat{n})\,,
\eea
where $\hat{n}$ is the neutron spin polarization axis.
We follow the convention for coordinates in \cite{rustgi60},
from which we have $\hat{k} = (0,\, 0,\, 1)$,
$\hat{n}= \hat{y}' = (-\sin\phi,\cos\phi,0)$,
and 
$\hat{p} = 
(\sin \theta \cos \phi,\, \sin \theta \sin \phi,\, \cos\theta)$,
where $\theta$ and $\phi$ are the colatitude and azimuthal angles 
in the c.m. frame.
Inserting the projection operator in the spin-isospin summation of
the squared amplitude, we obtain
\bea
S^{-1}\sum_{spin}^P |A|^2 &=&
4 \left[
|X_{MS}|^2
+|Y_{MV}|^2
-\left(X_{MS}^*Y_{MV}
+Y_{MV}^*X_{MS}\right)
\right] \nnb \\ &&
+2 \left[
|X_{MV}|^2
+|Y_{MS}|^2
- \left(
X_{MV}^* Y_{MS}
+Y_{MS}^*X_{MV}
\right)
\right]
\nnb \\ &&
+3\left[
1 - (\hat{k}\cdot \hat{p})^2
\right] \left[
|X_{EV}|^2
+|Y_{ES}|^2
-\left(
X_{EV}^*Y_{ES}
+Y_{ES}^* X_{EV}
\right)
\right]
\nnb \\ &&
\pm i \hat{n}\cdot (\hat{k}\times\hat{p})\left[
\left(X_{EV}^* X_{MV}
- X_{MV}^* X_{EV}
\right)
+ \left(Y_{ES}^* Y_{MS}
- Y_{MS}^*Y_{ES}
\right)
\right. \nnb \\ && \left.
-\left(
Y_{ES}^* X_{MV}
-X_{MV}^*Y_{ES}\right)
- \left(
Y_{MV}^*Y_{MS}
- Y_{MS}^*Y_{MV}
\right)
\right]\,.
\eea
Noting that, whereas $X_{MV}$ and $X_{MS}$ are complex, 
$X_{EV}$, $Y_{MV}$, $Y_{MS}$ and $Y_{ES}$ are real, 
we arrive at a final form for the polarization $P_{y'}$ as
\bea
P_{y'} &=& \frac{\sigma_+(\theta)-\sigma_-(\theta)}{
\sigma_+(\theta)+\sigma_-(\theta)}
\nnb \\ &=&
-2 \sin\theta (X_{EV}-Y_{ES})\, {\rm Im} X_{MV}/\left\{
4\left(
|X_{MS}|^2
+|Y_{MV}|^2
-2Y_{MV}\, {\rm Re} X_{MS}
\right)
\right. \nnb \\ && \left.
+ 2\left(
|X_{MV}|^2
+|Y_{MS}|^2
-2Y_{MS}\, {\rm Re} X_{MV}
\right)
+3(1-\cos^2\theta)\left(
|X_{EV}|^2
+|Y_{ES}|^2
-2X_{EV}Y_{ES}
\right)
\right\}\,.
\nnb \\
\eea
Since $\hat{x}' = (\cos\theta \cos\phi,\, \cos\theta \sin\phi,\, 
- \sin\theta)$ and 
$\hat{z}' = (\sin\theta \cos\phi,\, \sin\theta \sin\phi,\,
\cos\theta)$, one can easily verify that 
$P_{x'}$ and $P_{z'}$ vanish in the chosen coordinate system.

\section{Results and Discussion}
\label{section5}

\begin{figure}[tbp]
\begin{center}
\epsfig{file=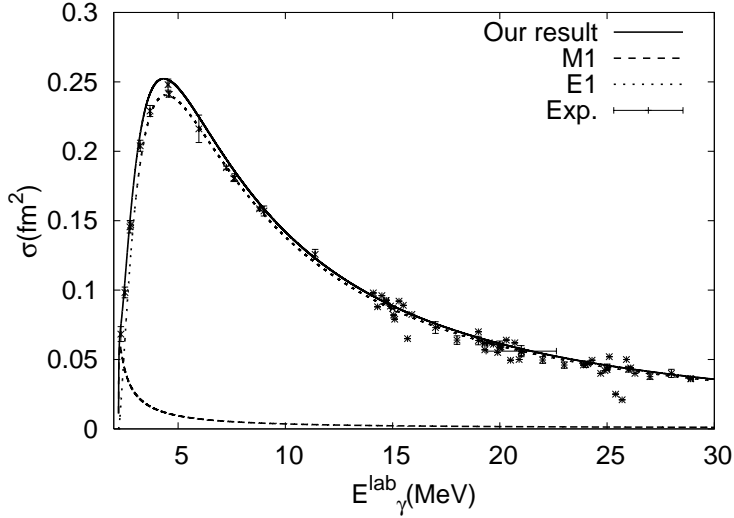,width=10.0cm}
\caption{\it 
Total cross section, $\sigma$,
for the $\gamma d\to np$ process
as a function of the lab-frame photon energy $E^{lab}_\gamma$ (MeV). 
The long-dashed and short-dashed lines 
show the M1 and E1 contributions, respectively;
the solid line gives the sum
of the E1 and M1 contributions. 
The points with error bars represent experimental data.
(The experimental data are obtained from 
National Nuclear Data Center (NNDC) web-page~\cite{nndc}.)
\label{fig;total-cross-section-dgnp}}
\end{center}
\end{figure}

Fig.~\ref{fig;total-cross-section-dgnp} 
shows the total cross section, $\sigma$,
for the $\gamma d \to np$ reaction from threshold
to $E^{lab}_\gamma =30$ MeV.
It is seen that, at low energies, 
there are a few data points 
that are off the calculated $\sigma$ curve. 
We remark, however,
that the error bars in the figure only represent statistical
errors, and that the error bars 
are likely to become significantly larger when systematic
errors are included.
The data at higher energies, $E_\gamma^{lab} \geq 15$ MeV 
exhibit some scatter, 
but their overall behavior is consistent 
with the calculated cross sections.
Thus we conclude that our dEFT calculation up to NLO, 
which contains no adjustable parameters after the two LECs 
($L_1$ and $L_2$) have been fixed,
can reproduce the total cross section data reasonably well.

\begin{figure}[tbp]
\begin{center}
\epsfig{file=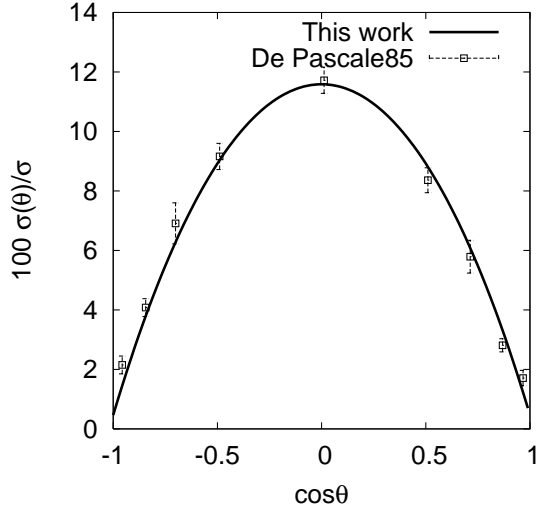,width=10.0cm}
\caption{\it
Differential cross section $100\, \sigma(\theta)/\sigma$ 
at $E^{lab}_\gamma = 19.8$ MeV,
where $\theta$ is the scattering angle in the c.m. frame.
The experimental data labeled ``De Pascale85" 
are taken from Ref.~\cite{pascale85}.
\label{fig;differential-cross-section-dgnp}}
\end{center}
\end{figure}

In Fig.~\ref{fig;differential-cross-section-dgnp} we plot the
differential cross section $\sigma(\theta)$ at
$E^{lab}_\gamma  = 19.8$ MeV, 
where $\theta$ is the scattering angle in the c.m. frame;
the figure also shows the data from \cite{pascale85}.
It can be seen that the calculated differential cross section
is consistent with the measurement.

\begin{figure}[tbp]
\begin{center}
\epsfig{file=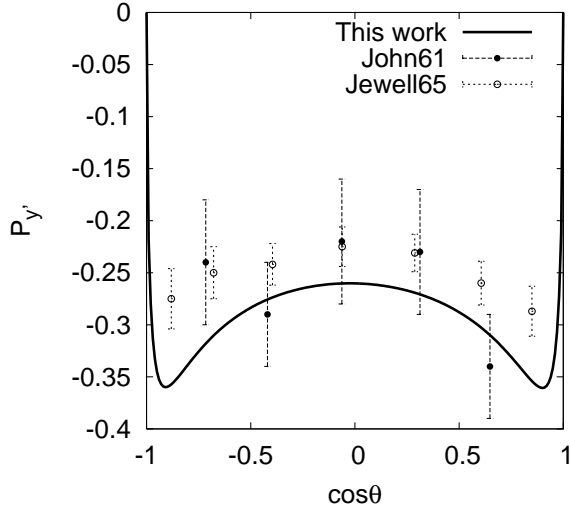,width=10.0cm}
\caption{\it
$P_{y'}$ at $E^{lab}_\gamma = 2.75$ MeV
plotted as a function of $\cos \theta$.
The experimental data labeled ``John61" are from \cite{john61},
and those labeled ``Jewell65" from \cite{jewell65}. 
\label{fig;pyp_y-dgnp}}
\end{center}
\end{figure}

Fig.~\ref{fig;pyp_y-dgnp} shows 
the angular distribution of the polarization $P_{y'}$
calculated for $E^{lab}_\gamma = 2.75$ MeV,
along with the experimental data.
There are two sets of data available in the literature.
One set (referred to as John61) is  
due to John {\it et al.}~\cite{john61},
and the other set (referred to as Jewell65)
is due to Jewell {\it et al.}~\cite{jewell65}.
We note that John61 has significantly
larger error bars than Jewell65.
Fig.~\ref{fig;pyp_y-dgnp} indicates 
that our results agree with John61 
within the large error bars
except at $\cos \theta \simeq -0.75$.
However, compared with Jewell65,
the theoretical curve clearly lies below 
the experimental values for the entire angular range.
In fact, this pattern of discrepancy 
between theory and experiment
was already discussed in Ref.~\cite{jewell65},
where the authors used the theoretical values of $P_{y'}$
that would turn out to be close to what we have obtained here. 
It is to be added that $P_{y'}$ for $E^{lab}_\gamma = 2.75$ MeV
calculated in SNPA~\cite{schi-prc05} agrees with our results.
Thus the data set Jewell65,
which has much smaller error bars than the earlier set John61,
disagrees with both the latest SNPA 
and dEFT calculations.\footnote{
We remark {\it en passant} that, 
if we multiply the calculated values of $P_{y'}$
with a factor of about 
0.7, 
the scaled results agree with Jewell65.}
%
%
The persistence of the $P_{y'}$ puzzle 
suggests the desirability of
a new measurement of $P_{y'}$. 

\begin{figure}[tbp]
\begin{center}
\epsfig{file=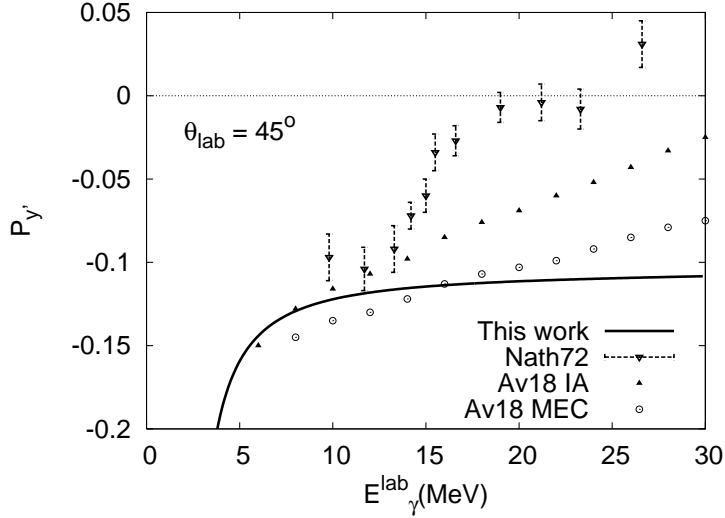,width=10.0cm}
\caption{\it
$P_{y'}$ at $\theta_{lab} = 45^\circ$ plotted
as a function of $E_\gamma^{lab}$,
where $\theta_{lab}$ is the scattering angle 
in the laboratory frame.
The data labeled ``Nath72" are from \cite{nath72}.
Filled triangles and open circles are 
results of IA and MEC in SNPA, respectively~\cite{schi-prc05}.
\label{fig;pyp45-dgnp}}
\end{center}
\end{figure}

Fig.~\ref{fig;pyp45-dgnp} 
shows, as a function of $E_\gamma^{lab}$,
the $P_{y'}$ for the lab-frame scattering angle 
$\theta_{lab} = 45^\circ$ calculated in dEFT;
also shown are the experimental data taken from \cite{nath72}. 
The figure includes the results of 
the previous SNPA calculation~\cite{schi-prc05} as well;
the values labeled ``IA" correspond to the impulse approximation (IA)
while those labeled ``MEC" include 
the meson exchange currents.
We can see from Fig.~\ref{fig;pyp45-dgnp} 
that the results of our dEFT calculation
completely disagree with the data.
The figure also indicates that the present dEFT
calculation gives 
values of $P_{y'}$
significantly different 
from those obtained in the SNPA calculation~ \cite{schi-prc05}
(although, for $E^{lab}_\gamma \leq 10$ MeV,
the dEFT curve happens to be rather close 
to the IA values in \cite{schi-prc05}.)
Whereas the SNPA values increase almost linearly
as functions of $E_\gamma^{lab}$
(with the IA curve increasing more rapidly than the MEC curve), 
$P_{y'}$ obtained in dEFT almost tapers off 
around $E_\gamma^{lab}\approx 10$ MeV.
This latter feature worsens the discrepancy 
between experiment and theory,
which was already conspicuous with the use of
the SNPA values of $P_{y'}$.

\begin{figure}[tbp]
\begin{center}
\epsfig{file=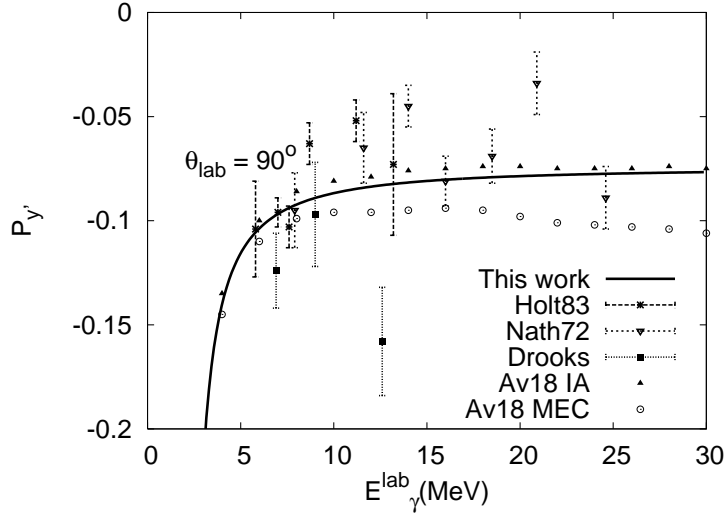,width=10.0cm}
\caption{\it
$P_{y'}$ at $\theta_{lab} = 90^\circ$
plotted as a function of $E^{lab}_\gamma$.
The data labeled ``Holt83" and ``Drooks" are from
\cite{holt83} and \cite{drooks}, respectively.
Filled triangles and open circles are 
results of IA and MEC in SNPA, respectively~\cite{schi-prc05}.
\label{fig;pyp90-dgnp}}
\end{center}
\end{figure}

Fig.~\ref{fig;pyp90-dgnp} 
shows, as a function of $E_\gamma^{lab}$,
the $P_{y'}$ at $\theta_{lab} = 90^\circ$ calculated in dEFT,
together with three sets of experimental data 
taken from Refs.~\cite{nath72,drooks,holt83}.
Comparison between theory and experiment
is hampered by the fact 
that the data points show pronounced scatter,
exhibiting at some places even inconsistencies
among the data sets from the different sources. 
It is also curious that, 
for $E^{lab}_\gamma \geq 10$ MeV, 
the data points show rather conspicuous oscillatory behavior.
We may summarize the current situation
with the statement that the average behavior
of the experimental values of $P_{y'}(\theta_{lab} \!\!= 90^\circ)$ 
agrees with the results of our dEFT calculation
within the very large experimental uncertainties.
Here again, it seems desirable 
to have new measurements of $P_{y'}$.
Fig.~\ref{fig;pyp90-dgnp} also gives $P_{y'}$
obtained in a SNPA calculation~\cite{schi-prc05}.
It is seen that, within SNPA, 
the IA calculation always gives larger values 
of $P_{y'}$ than the MEC case.
The curve corresponding to our dEFT calculation 
lies between the IA and MEC values 
for $E^{lab}_\gamma \leq 10$ MeV,
but it approaches the IA results as $E^{lab}_\gamma$ increases.

\begin{figure}[tbp]
\begin{center}
\epsfig{file=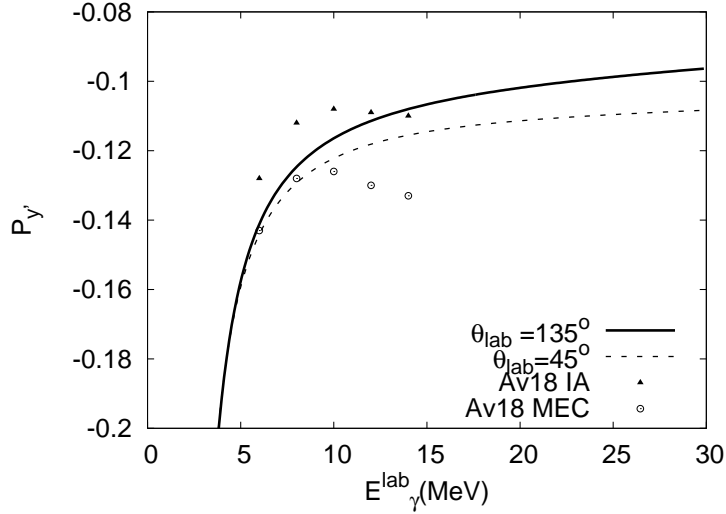,width=10.0cm}
\caption{\it
$P_{y'}$ plotted as a function of $E^{lab}_\gamma$
for $\theta_{lab} = 135^\circ$ (solid line)
and $\theta_{lab} = 45^\circ$ (dashed line).
Filled triangles and open circles are 
results of IA and MEC in SNPA, respectively~\cite{schi-prc05}.
\label{fig;pyp135-dgnp}}
\end{center}
\end{figure}

In Fig.~\ref{fig;pyp135-dgnp}, we plot the dEFT values of 
$P_{y'}$ at $\theta_{lab} = 135^\circ$, 
as a function of $E_\gamma^{lab}$.
For comparison, we also show the SNPA values of $P_{y'}$
for the IA and MEC cases~\cite{schi-prc05}.
It is to be seen 
that, for $E^{lab}_\gamma \leq 5$ MeV,
the dEFT and SNPA results are close to each other,
but that qualitative difference appears 
for $E^{lab}_\gamma \geq 8$ MeV;
the dEFT curve  shows slow, monotonic increase, 
whereas the SNPA results (both IA and MEC cases)
start decreasing  around $E^{lab}_\gamma \geq 10$ MeV.
We also remark that, in our dEFT calculation,
the dominant contributions to $P_{y'}$
are proportional to $\sin \theta$ or $\cos^2\theta$,
which implies $P_{y'}(\theta_{lab} \!\!= 135^\circ)\approx 
 P_{y'}(\theta_{lab} \!\!= 45^\circ)$.
Fig.~\ref{fig;pyp135-dgnp}, which also includes 
 $P_{y'}(\theta_{lab} \!\!= 45^\circ)$,
indicates  that 
$P_{y'}(\theta_{lab} \!\!= 135^\circ)\approx 
 P_{y'}(\theta_{lab} \!\!= 45^\circ)$ holds rather well.
By contrast, the calculation in \cite{schi-prc05}
does not share this feature,
another qualitative difference between our results 
and those in \cite{schi-prc05}.
%
%
%
%
As regards comparison with experiment,
because we were unable to retrieve the relevant data from the literature,
we cannot make direct comparison
of the theoretical $P_{y'}(\theta_{lab} \!\!= 135^\circ)$ with experiment.
However, to the extent that $P_{y'}(\theta_{lab} \!\!= 135^\circ)\approx 
 P_{y'}(\theta_{lab} \!\!= 45^\circ)$ holds,
we can compare our theoretical curve  with the data
for $P_{y'}(\theta_{lab} \!\!= 45^\circ)$ 
shown in Fig.~\ref{fig;pyp45-dgnp} .

\section{Conclusions}
\label{section6}

We have applied the pionless-EFT-with-dibaryon (dEFT) formalism
to the  $\gamma d$ $\to$ $\vec{n} p$ reaction 
for the incident photon energy up to $30$ MeV. 
As far as the total cross section and the differential cross section
are concerned, the results of our dEFT calculation
agree with those of the latest SNPA (standard nuclear physics approach)
calculation by Schiavilla~\cite{schi-prc05}.
These theoretical values are in reasonable agreement with the data,
which at present have appreciable uncertainties.
On the other hand, for the neutron polarization $P_{y'}$, 
the results of our dEFT calculation are found to be 
significantly different from those obtained in SNPA~\cite{schi-prc05}, 
indicating the sensitivity of polarization observables
to the theoretical frameworks used.
It is noteworthy that,
even if we interpret the difference between the EFT and SNPA results
as a rough measure of the existing theoretical uncertainties,
the  ``$P_{y'}$" puzzle,
{\it i.e.}, discrepancy between the theoretical and experimental values of  ``$P_{y'}$",
still persists.  
Our results indicate that  $P_{y'}$ obtained in dEFT 
can exhibit even larger discrepancy
from the data than the SNPA calculation does,
for certain ranges of the scattering angle.

%
%
We remark that, at energies larger than $E_\gamma^{lab}\sim 15$ MeV,
the contributions of final-state partial waves higher than considered 
in our present calculation may become significant, and that
the SNPA calculation~\cite{schi-prc05} includes these higher partial waves.
This may explain part of the differences between the dEFT and SNPA 
results for $P_{y'}$.   
Higher order effects within dEFT also need to be examined
despite the good convergence property of dEFT found previously
for many observables.
It is to be noted, however, that
Christlmeier and Grie{\ss}hammer~\cite{CHRetal}
have carried out an N$^2$LO calculation in dEFT
for the longitudinal and transverse response functions 
for the $d(e,e')$ reaction.  
According to this work, it is highly unlikely 
that the large discrepancy between 
theory and experiment found for some of these response functions
can be ascribed to higher order terms
in the dEFT expansion.
A similar conclusion may hold for $P_{y'}$,
and a calculation going up to N$^2$LO
for the $\gamma d$ $\to$ $\vec{n} p$ reaction
seems warranted.
We also remark that,
even at energies below 10~MeV, 
the inclusion of higher order corrections is desirable 
in that it will reduce theoretical uncertainties
and help sharpen the issue of the discrepancy 
between dEFT and SNPA at low energies.

We make here a brief comment 
on the treatment of  the internal structure of the deuteron in dEFT.  
The introduction of the elementary dibaryons,
$d_s$ and $d_t$, in dEFT might give the impression 
that the deuteron structure has no place in dEFT. 
It is to be noted, however, that 
a photon coupled to the intermediate nucleon (diagram (b) in Fig. 1) 
gives rise to momentum dependence in the deuteron form factor,
and thus the structure effects subsumed in the form factor 
can be accommodated in dEFT.
In Ref.~\cite{ando-prc05}, 
the electromagnetic form factors for the
deuteron were calculated in dEFT up to N$^3$LO, and the differential
cross sections for $e$-$d$ elastic scattering were computed
with the use of these form factors.
The functions, $A(q)$ and $B(q)$, that represent the momentum
dependence of the cross section 
(see Ref.~\cite{ando-prc05} for the definitions of these functions)
were compared with the experimental data and 
also with the results of other theories,
and good agreement in the low-momentum transfer region
was reported. 
Given the generally good convergence properties of dEFT,
we may expect that our present NLO calculation
incorporates the bulk of the deuteron structure effects,
even though a possibility does exist that $P_{y'}$ is a 
``delicate" quantity that is exceptionally sensitive to 
higher-order terms.
In this context also, an extension of the present work to 
higher chiral orders seems of importance.

It is worth emphasizing that
the accurate understanding of polarization observables 
is also important in connection with parity-violating
observables in nuclear electromagnetic processes \cite{bertrand98}. 
In the process $\gamma d\to\vec{n}p$, for example,
the neutron polarizations along the 
$\hat{x}'$ and $\hat{z}'$ directions vanish with
the parity-conserving interactions as mentioned before, 
but they can be non-vanishing with the parity-violating interactions.
Theoretical prediction on these parity-violation observables
requires high accuracies in both the 
strong and electromagnetic amplitudes.
A polarization observable which is sensitive to the interference
between the strong and the electromagnetic amplitudes
can be a good testing ground for the
reliability of parity-violation calculations
as well.

To summarize, our study points to the necessity of further studies,
both experimental and theoretical, 
of the spin observables in the  $\gamma d$ $\to$ $np$ reaction.

\section*{Acknowledgments}

The work of SIA and CHH is supported by the Basic Science Research 
Program through the National Research Foundation of Korea (NRF)
funded by the Ministry of Education, Science and Technology 
(2010-0023661).
The work of YHS is supported by the US Department of Energy 
under Contract No. DE-FG02-09ER41621.
KK's work is partly supported by the US National Science Foundation
under grant number PHY-0758114.

\end{document}